\documentclass[aps,prb,twocolumn]{revtex4}
\input epsf
\usepackage{graphicx}

\begin{document}
\title{Atomic step motion during the dewetting of ultra-thin films}
\author{O. Pierre-Louis$^{1,2}$, A. Chame$^3$, M. Dufay$^{1,2}$}
\affiliation{$^1$ LPMCN, Universit\'e  Lyon 1, 43 Bd du 11 novembre, 69622 Villeurbane, France.\\
$^2$Laboratoire de Spectrom\'etrie Physique,
UJF Grenoble 1, BP 87, 38402 St Martin d'H\`eres, France.\\ 
$^3$ Universidade Federal Fluminense,Avenida Litor\^anea s/n, 24210-340 Niter\'oi RJ, Brazil
}
\date{\today}

\begin{abstract}
We report on three key processes involving atomic step motion
during the dewetting of thin solid films: (i) the growth 
of an isolated island nucleated far from a hole, (ii) the
spreading of a monolayer rim, and (iii) the zipping of
a monolayer island along a straight dewetting front.
Kinetic Monte Carlo results are in good agreement
with simple analytical models assuming diffusion-limited dynamics.
\end{abstract}

\maketitle

\newpage
\section{Introduction}

The shape changes of crystals are usually governed by
the motion of atomic steps at the surface\cite{Burton1951,Misbah2010}. 
The study of steps dynamics have
permitted a better understanding of the morphological changes during various 
non-equilibrium processes, such as crystal growth\cite{Misbah2010},
surface electromigration\cite{Misbah2010},
or the decay of performed structures\cite{Chame1996}. In the present paper, we focus
on some key processes involving the dynamics of 
atomic steps during the dewetting of ultra-thin solid films.
Dewetting is the process by which a continuous
film breaks down into islands to lower the global energy.
A large body of theoretical work has been devoted to the dynamics of dewetting
with various continuum models\cite{Khenner2008b,Khenner2008,Kan2005,Wong2000,Aqua2007,Srolovitz1986}.
Nevertheless, Kinetic Monte Carlo (KMC) simulations\cite{PierreLouis2007,PierreLouis2009}
have revealed that the dewetting process is controlled
by the nucleation and dynamics of atomic steps on facets.
Furthermore, recent studies have reported the observation
of facets in dewetting experiments with Ag/Si(111) \cite{Thurmer2003},
SOI (Si/Si0$_2$) systems \cite{lagally05,Dornel2006},
and YBaCuO films \cite{Coll2006},
hence suggesting that atomic steps could play a crucial role
in the dynamics.

A full analysis of the morphological evolution during dewetting
would require to investigate not only step motion, but also the 
formation of new steps via two-dimensional nucleation.
Such an approach was proposed recently in Refs.\cite{PierreLouis2007,PierreLouis2009}.
However, a quantitative description of nucleation is in general delicate because it
requires an accurate determination of the step free
energy which enters in the nucleation barrier.
We shall here discard the nucleation process, and focus on the
motion of steps in various geometries.

The paper is organized as follows.
In section \ref{s:kmc}, we provide a description of the KMC model.
Section \ref{s:island_and_hole} focuses on the layer-by-layer dewetting regime.
In this regime, an island is first nucleated far from a
pre-existing hole in the film. The island then grows by
diffusion-limited mass transfer from the hole. We shall here present
an analytical solution of the island growth process,
which provides a good agreement with the 
KMC simulations. The island finally grows so much
that it surrounds the hole, and forms a monolayer 'rim' around it.
In section \ref{s:monolayer_rim}, we provide an expression 
for the dynamics of the monolayer rim
once it has surrounded the hole. This expression
is in quantitative agreement with the simulations, and 
contains some logarithmic corrections as compared
to the simple linear behavior proposed in Ref\cite{PierreLouis2009}.
In section \ref{s:zipping}, we consider a different regime
where a thick facetted rim forms during the dewetting process.
Islands nucleated on the rim facet then grow in size
and spread along the dewetting front with a well defined
velocity $V_{zip}$. The numerical evaluation of this velocity
is difficult because it would require delicate front-tracking procedures.
We therefore present a method which allows one to determine this velocity
from a single snapshot of KMC simulations. This method,
the results of which were presented in Ref.\cite{PierreLouis2009}, is
reported here in details.

\section{KMC simulations}
\label{s:kmc}

We model the dewetting of a crystalline film  using a
solid-on-solid model on  a 2D square  lattice.
The lattice unit is $a$.
The substrate surface is perfectly flat and frozen.
The local height is $z$. On the epilayer atoms $z>0$, and on
the substrate $z=0$.

\begin{figure}
\includegraphics[height=3 cm]{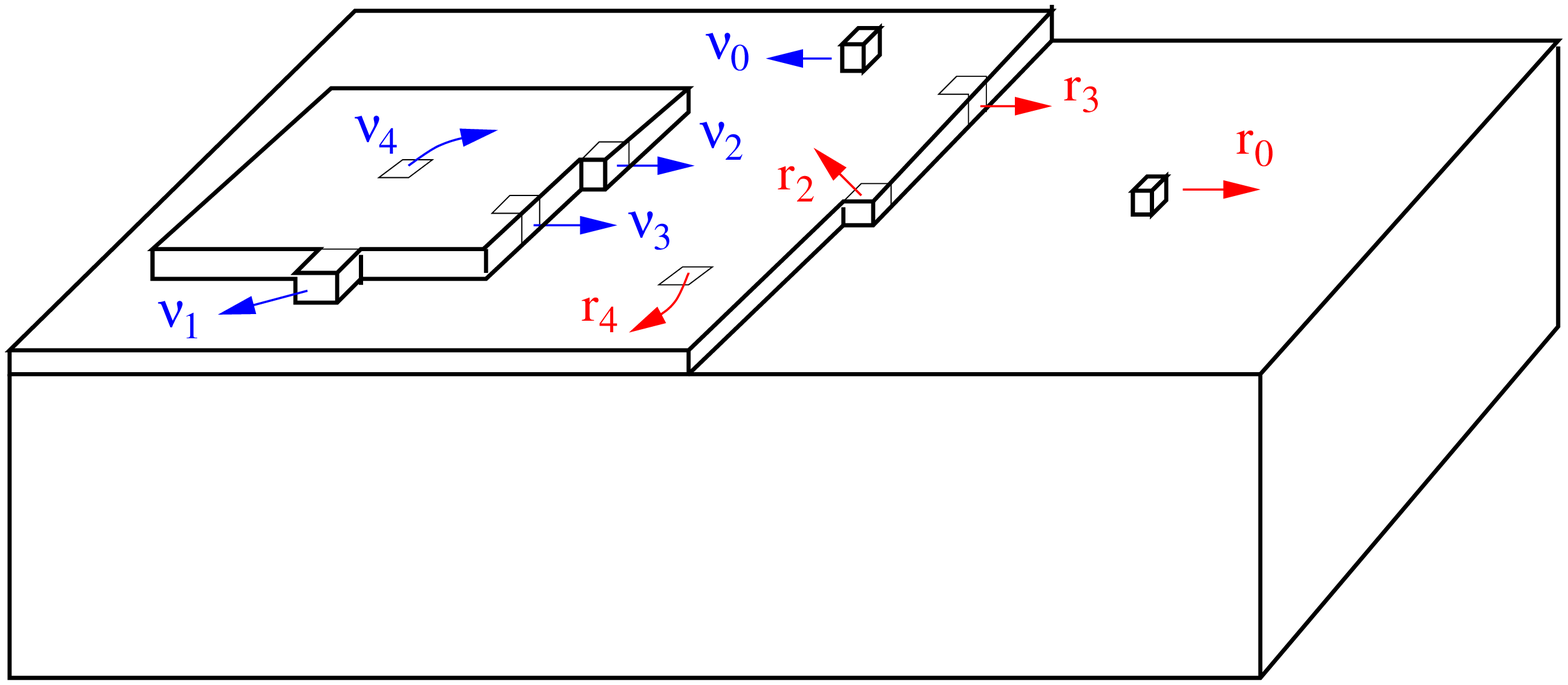}\\
\caption{Kinetic Monte Carlo (KMC) model, with hopping rates $r_n$ and $\nu_n$.}
\label{fig:KMC_model}
\end{figure}

We employ KMC simulations
to implement the dynamics.  Epilayer atoms hop to nearest neighbor
sites with rates $r_n$ when they are in contact with
the substrate ($z=1$), and $\nu_n$ when they are not in direct
contact with the substrate ($z>1$). In our model, an atom needs to
break all its bonds to hop.
The hopping barrier is therefore given by the
binding energy of the atom. Hence:
\begin{eqnarray}
r_n &=& \nu_0 {\rm e}^{-nJ/T + E_S/T}
\label{e:r_n}
\\
\nu_n &=& \nu_0 {\rm e}^{-nJ/T}
\label{e:nu_n}
\end{eqnarray}
where $\nu_0$ is an attempt frequency,
$T$ is the temperature (in units with $k_B=1$),
$n$ is the number of in-plane nearest neighbors of the 
atom before the hop,
$J$ is the bond energy, and $E_S$ is the adsorbate-substrate
excess energy. The model is presented in Fig\ref{fig:KMC_model}. 
Note that we do not write explicitly the
bond $J$ between the moving atom and the atom
directly underneath it in the energy barrier for
atom motion in Eqs.(\ref{e:r_n},\ref{e:nu_n}).
Indeed, since this bond is present below any atom
before the hopping event, we rather include
its contribution $\exp[-J/T]$ in the prefactor $\nu_0$.
We choose $J$ as the energy unit, 
so that $J=1$ in the following.

The algorithm used in the simulations is the following.
We  list all atoms into  classes. Each class
is characterized by the number of  in-plane  neighbors $n$ of the atom and by the
existence or not of  a nearest neighbor  belonging to the substrate. At a given time t,
we calculate the probabilities per unit time $w_i$ of all possible events ( an event is
the motion of an atom originally at position $i$ ) given either by Eq.(\ref{e:r_n}) or
Eq.(\ref{e:nu_n}), and the sum W of all those rates (for all mobile atoms).
We increment the time by a $\delta t$, which is equal to the inverse
of the sum of the rates of all possible events, $1/W$. This choice for
$\delta t$  corresponds to  the average value of the waiting time between
two successive events (Poisson processes)  \cite{Kotrla1996}.
We choose the event with probability $w_i/W$. The chosen atom
moves with equal probability in any of the four possible  directions.

The parameter $E_S$ controls the wetting properties
of the film on the substrate. 
In order to relate more precisely $E_S$ to surface
energetics, let us perform two Gedankenexperiments.
The first one consist in splitting an epilayer crystal
into two parts perpendicular to the $z$ axis.
We obtain two surfaces, and the energy balance
is: $J=2E_{AV}$, where $E_{AV}$ is the energy
per site of the created facet.
The second Gedankenexperiment is to
separate the substrate and the epilayer.
The energy balance now reads: $E_{AS}+J-E_S=E_{AV}+E_{SV}$,
where $E_{AS}$ and $E_{SV}$ is the epilayer-substrate 
and substrate-vacuum energies per lattice site.
Combining these two relations, we find 
\begin{eqnarray}
E_S=E_{AV}+E_{AS}-E_{SV}.
\end{eqnarray}
Since the simulations were performed at a fixed temperature $T$,
we should in principle consider a balance between free energies,
rather than energies. However, since the substrate is frozen,
the epilayer-substrate, and substrate-vacuum interfaces
do not exhibit any configurational entropy. Their energy is therefore
equal to their free energy. In addition we will work below
the roughening temperature, so that the epilayer-vacuum free energy
can also be approximated by an energy.
As a consequence, the parameter $E_S$ is, to a good approximation,
a balance of free energies. This parameter is
then identical (with opposite sign) to the thermodynamic work
of adhesion $S$
defined in Ref.\cite{DeGennes1985,SanGiorgi1988}.

When $E_S\leq 0$, we have complete wetting.
The regime of partial wetting is obtained for $E_S>0$.
At low temperatures, the equilibrium island
shape has  a square base of lateral size $L_{eq}$,
and a height $h_{eq}$, with $h_{eq}/L_{eq}=E_S$.
When $E_S\gg1$, the energy minimization will
favor high islands. As $E_S$ is decreased
to lower values, the equilibrium aspect ratio of the island decreases,
and the islands become flat for $E_S\ll 1$.

\section{Island and hole geometry}
\label{s:island_and_hole}
\subsection{Model}

\begin{figure}
\includegraphics[height=11 cm]{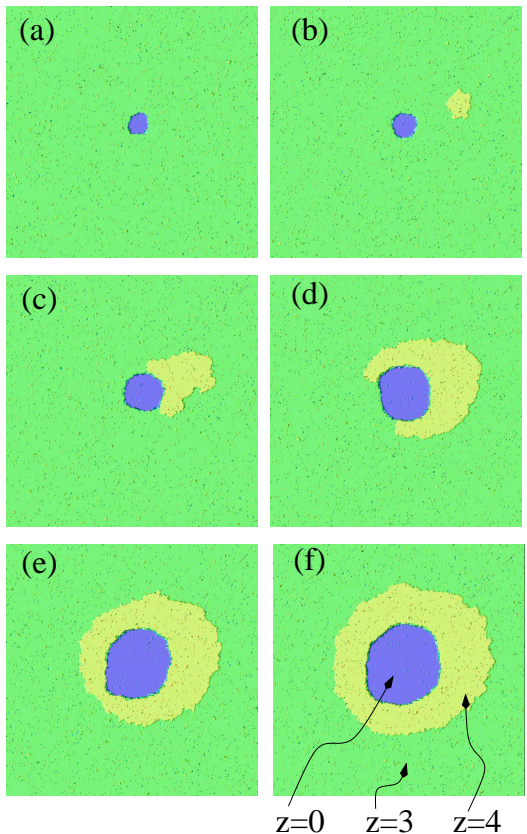}\\
\caption{Typical dynamics in the layer-by-layer dewetting regime. Parameters:
$T=0.4$, $E_S=0.25$, system size $400\times400$, and $h=3$. (a) A hole is performed
artificially in the film. (b) An island nucleates far from the hole.
(c) Diffusion limited mass transfer from the hole to the island
leads to the simultaneous growth of the island and the hole.
(d) The monolayer island surrounds the hole.
(e) The monolayer island closes around the hole and forms a monolayer rim.
(f) The monolayer rim grows and becomes circular.}
\label{fig:island_and_hole}
\end{figure}

Let us consider the dewetting process starting from
a pre-existing hole in a film of thickness $h$. Such a process 
mimics dewetting initiated at heterogeneous nucleation sites.
When $E_S$ is small or $h$ is large, dewetting occurs
in a layer-by-layer fashion.The precise criterion for this regime
to occur was given in Ref.\cite{PierreLouis2009}. 
The process starts with the nucleation
of a monolayer island on the film.
Since this nucleation event is randomly located\cite{PierreLouis2009},
the island hole distance is usually large,
as shown in Fig.\ref{fig:island_and_hole}(b).
The island and the hole then grow simultaneously
due to mass transfer from the hole to the island.
We shall here focus on this island and hole
growth process, and propose an analytical
expression for the dynamics.
We assume that the hole-island pair is
isolated on the surface: no other hole or island
exists in its neighborhood.
In the initial stages, the distance between the hole and the island
is larger than the typical sizes of the hole and the island.
In this limit, we may assume that the hole and the island
are circular. We also assume that the dynamics is diffusion-limited,
as in Ref.\cite{PierreLouis2009}.
From the assumption of fast adatom attachment-detachment kinetics
at the island and hole edges, the concentrations is fixed to its equilibrium
value. Hence, the concentrations is $c_{eq}^{A}$ at the hole edge,
and $c_{eq}^B$ at the island edge.
In such a geometry, the diffusion problem can be solved 
in bi-polar coordinates \cite{Morse1953}.
The adatom mass flux from the hole to the island is then found to be:
\begin{eqnarray}
J={2\pi D \Delta c\over \ln[\pi d^2/(AB)^{1/2}]},
\end{eqnarray}
where $\Delta c=c_{eq}^{A}-c_{eq}^B$, and 
$A$ and $B$ are the areas of the hole and of the island,
and $d$ is the distance between the center of the hole and 
the center of the island.
From mass conservation, the flux $J$ is then related to the 
evolution of the hole area: 
\begin{eqnarray}
\partial_tA={\Omega \over h}J={2\pi \Omega D\Delta c \over h \ln[d^2\pi/(AB)^{1/2}]},
\label{e:Bevol1}
\end{eqnarray}
where $\Omega=a^2$ is the atomic area.
The hole and island areas are then related via global mass conservation
\begin{eqnarray}
\partial_tB=h\partial_tA.
\label{a:mass_cons_island_hole}
\end{eqnarray}
Since the island forms after the hole, we may
assume that at $t=0$, $A=A_0$ and $B=0$,
so that 
\begin{eqnarray}
A={B \over h}+A_0.
\end{eqnarray}
Using this relation into Eq.(\ref{e:Bevol1}),
we find:
\begin{eqnarray}
\partial_tA= {2\pi\Omega D\Delta c \over h \ln[d^2\pi/(hA(A-A_0))^{1/2}]}
\end{eqnarray}
the solution of which can be written in the implicit form:
\begin{eqnarray}
\bar A_0\ln\left[\bar A_0\over \bar A\right]-(\bar A-\bar A_0)
\ln\left[{h\over {\rm e}^{2}}\bar A(\bar A-\bar A_0)\right]= {4\Omega D\Delta c \over hd^2}t,
\nonumber\\
\label{e:sol_island_hole}
\end{eqnarray}
where $\bar A=A/\pi d^2$, and $\bar A_0=A_0/\pi d^2$.
At long times, but when the hole is still small as compared
to the hole-island distance, i.e. when $1\gg\bar A\gg\bar A_0$, one has:
\begin{eqnarray}
\bar A \ln\left[ {\rm e}^2 \over \bar A^2h \right]
\approx {4\Omega D\Delta c \over hd^2}\, t.
\label{e:sol_island_hole_asympt}
\end{eqnarray}

\subsection{Comparison to KMC simulations}

In order to compare Eq.(\ref{e:sol_island_hole}) with KMC simulations,
we need to evaluate the parameters which enter in the model.
We choose the lattice spacing as our unit length, and the time unit is 
the inverse of the attempt frequency. Hence $\Omega=1$ and $\nu_0=1$. 
Since an atom can move randomly to the 4 nearest neighbor sites, 
the diffusion constant on the film is $D=1/4$. Moreover, 
we obtain the equilibrium concentration
on the film $c_{eq}={\rm e}^{-2J/T}$
from the detailed balance of attachment-detachment
at kink sites. Since the curvature
of the island is larger than the curvature of the critical
island size, we may assume that
the Gibbs-Thomson contributions proportional to the curvature,
are negligible, and $c_{eq}^B\approx c_{eq}$.

\begin{figure}
\includegraphics[height=5 cm]{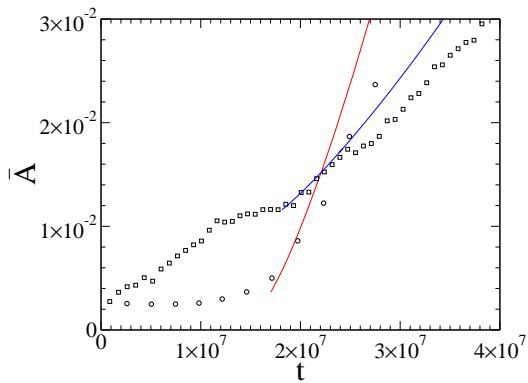}\\
\caption{Dynamics of one hole exchanging mass with a distant monolayer island.
The symbols are the results of KMC simulations with:
$T,E_S,h=0.4,0.2,2$ (squares), and $0.35,0.25,2$ (circles), in a $400\times400$ periodic lattice.
(In order to fit both curves in the same plot, the timescale of the first data set (squares) was multiplied by 6).
The solid lines represents the solution of Eq.(\ref{e:sol_island_hole}) 
where the initial conditions and the value of $d$ were measured from the simulations.}
\label{fig:area_norim_DL}
\end{figure}

The chemical potential in the vicinity of the film edge
is $E_S/Th$, as shown in Appendix \ref{a:chem_pot_edge_hole}.
Our reference sate is the state of equilibrium
on top of the film, with equilibrium concentration $c_{eq}$
(this is the equilibrium concentration in the vicinity of an
atomic step on the top of the film).
Using the standard formula:
\begin{eqnarray}
\mu=T\ln{c_{eq}^A\over c_{eq}},
\end{eqnarray}
we find that
\begin{eqnarray}
c_{eq}^A=c_{eq}\,{\rm e}^{\mu/T}
=c_{eq}\,{\rm e}^{E_S/Th}.
\end{eqnarray}
We therefore have
\begin{eqnarray}
\Delta c &=& {\rm e}^{-2J/T}\left({\rm e}^{E_S/(hT)}-1\right).
\label{e:param}
\end{eqnarray}
Using these parameters, the solution of Eq.(\ref{e:sol_island_hole}) is in good
agreement with KMC simulations, as shown in Fig.\ref{fig:area_norim_DL}.
Note that the Eq.(\ref{e:sol_island_hole}) cannot lead
to a perfect agreement with KMC simulations due to inherent assumptions
of the model. Indeed, we have assumed that the island and hole are
isolated, neglecting interaction with the periodic images
of the island and the hole. In addition, the condition
that the island-hole distance $d$ is much larger
that the radii  of the hole and the island is  verified at the beginning,
but not at end of the KMC simulation.
Finally, we observe in Fig.\ref{fig:area_norim_DL} that
the condition $A\gg A_0$ is not verified in the KMC simulations.
Therefore, the asymptotic linear solution (\ref{e:sol_island_hole_asympt}) 
cannot provide a good approximation for the KMC simulations result.

\section{Growth of a monolayer rim around a hole}
\label{s:monolayer_rim}

\subsection{Model}
Due to the smaller thickness  of the monolayer as compared to
that of the hole, the island area increases
faster than that of the hole, as shown by Eq.(\ref{a:mass_cons_island_hole}). 
Once the island becomes 
large enough, it collides with the hole and surrounds it.
Then, a different regime can be observed, where the 
hole is surrounded by a monolayer rim. This monolayer
rim is initially very far from being circular. 
However, the diffusion limited transport provides
more mass to the parts of the monolayer rim which are closer 
to the hole, resulting in a stabilization of the rim,
which becomes approximately circular.  In this section,
we present the exact solution of the circular rim
evolution dynamics, and compare this solution
to the asymptotic solution given in Ref.\cite{PierreLouis2009}.

Mass exchange  occurs on top of the monolayer rim via adatom diffusion.
The adatom concentration field on the rim obeys:
\begin{eqnarray}
{1 \over r}\partial_r\left[r\partial_r c\right]=0,
\end{eqnarray}
which has a solution of the form
\begin{eqnarray}
c=c_0\ln[r/r_0].
\end{eqnarray}
We once again assume instantaneous attachment-detachment
kinetics, so that $c=c_1$ at $r=R_1$ at the edge of the 
hole and $c=c_2$ at $r=R_2$ at the other edge.
Moreover, we must impose mass conservation
\begin{eqnarray}
\partial_tR_2&=&-{\Omega \over h_2} D\partial_rc|_2,
 \nonumber \\
\partial_tR_1&=&-{\Omega\over h_1} D\partial_rc|_1,
\end{eqnarray}
where $h_1=h+1$ is the rim height, and $h_2=1$ is the thickness
of the monolayer rim.
Introducing the areas $A_1=\pi R_1^2$ and $A_2=\pi R_2^2$,
we find 
\begin{eqnarray}
\partial_tA_1={4\pi D \Omega \Delta c \over
h_1 \ln[A_2/A_1]},
\label{e:A1evol1}
\end{eqnarray}
where $\Delta c=c_1-c_2$ is given by Eq.(\ref{e:param}) with the substitution
$h\rightarrow h_1$.
The evolution of $A_2$ is fixed by the mass conservation relation
\begin{eqnarray}
h_2\partial_tA_2=h_1\partial_tA_1,
\end{eqnarray}
which implies
\begin{eqnarray}
A_2={h_1\over h_2} (A_1-A_1^0)+A_2^0,
\end{eqnarray}
where $A_i^0$
are the areas at $t=0$.
Substituting this expression for $A_2$
into Eq.(\ref{e:A1evol1}), one finds the solution
in an implicit form:
\begin{eqnarray}
&&h_2(A_2\ln[A_2]-A_2^0\ln[A_2^0])
\nonumber \\
&&-h_1(A_1\ln[A_1]-A_1^0\ln[A_1^0])
\nonumber \\
&=&4\pi D\Omega\Delta c t.
\label{e:A1_evol_h1_cst}
\end{eqnarray}
At large times when $A_1\gg A_1^0$ and $A_2\gg A_2^0$, we find
\begin{eqnarray}
A_1h_1\ln\left[h_1\over h_2\right]
+A_1^0h_1\ln\left[A_1^0h_2\over{\rm e}A_1h_1\right]
-A_2^0h_2\ln\left[A_2^0h_2\over{\rm e}A_1h_1\right]
\nonumber \\
\approx 4\pi D\Omega\Delta c \; t.
\nonumber \\
\label{e:area_rim_mono}
\end{eqnarray}
Keeping the dominant terms only, we obtain a  linear behavior at large times
\begin{eqnarray}
A_1\approx {4\pi D\Omega\Delta c \over h_1\ln[h_1/h_2]}t
\label{e:area_rim_mono_asympt}
\end{eqnarray}
which is the asymptotic form presented
in Ref.\cite{PierreLouis2009}. 

\subsection{Comparison to KMC simulations}

\begin{figure}
\includegraphics[height=6 cm]{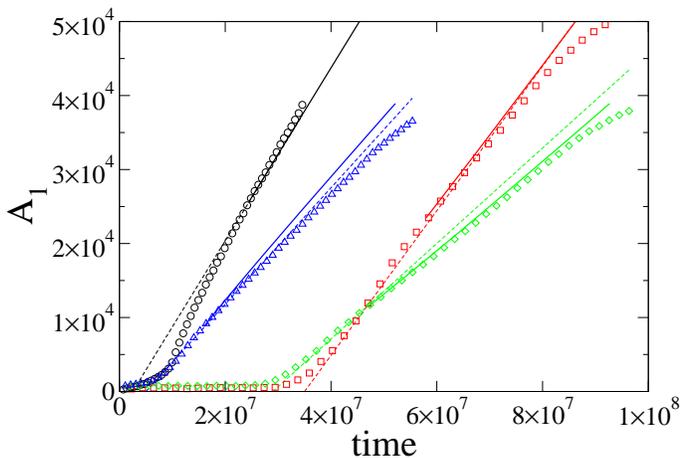}\\
\caption{Diffusion limited regime. Dynamics of $A_1$ for a
hole  with a monolayer rim. 
The dashed lines correspond to the asymptotic linear formula Eq.(\ref{e:area_rim_mono_asympt}).
The solid lines correspond to the full solution Eq.(\ref{e:A1_evol_h1_cst})
without fitting parameter.}
\label{fig:area_rim_mono}
\end{figure}

In Fig\ref{fig:area_rim_mono}, we have compared
the KMC simulations results  with Eq.(\ref{e:A1_evol_h1_cst}) and Eq.(\ref{e:area_rim_mono_asympt}).
The agreement between the linear asymptotic formula
(\ref{e:area_rim_mono_asympt}) and KMC simulations results
requires a fitting parameter
(e.g. the reference time for which $A_1^0$ extrapolates to zero).
However, due to the divergence of the subdominant terms in Eq.(\ref{e:area_rim_mono}),
the meaning of this parameter is unclear.

Therefore, it is more satisfactory to use the full solution (\ref{e:A1_evol_h1_cst}).
Indeed, the full solution requires no fitting parameter: we simply need to measure $A_1^0$ and
$A_2^0$ at an initial time once the monolayer rim is formed.
The full solution is shown to be in good agreement
with the KMC simulations in Fig.\ref{fig:area_rim_mono}.

\section{Zipping of a monolayer on a facetted rim}
\label{s:zipping}

\begin{figure}
\includegraphics[height=13 cm]{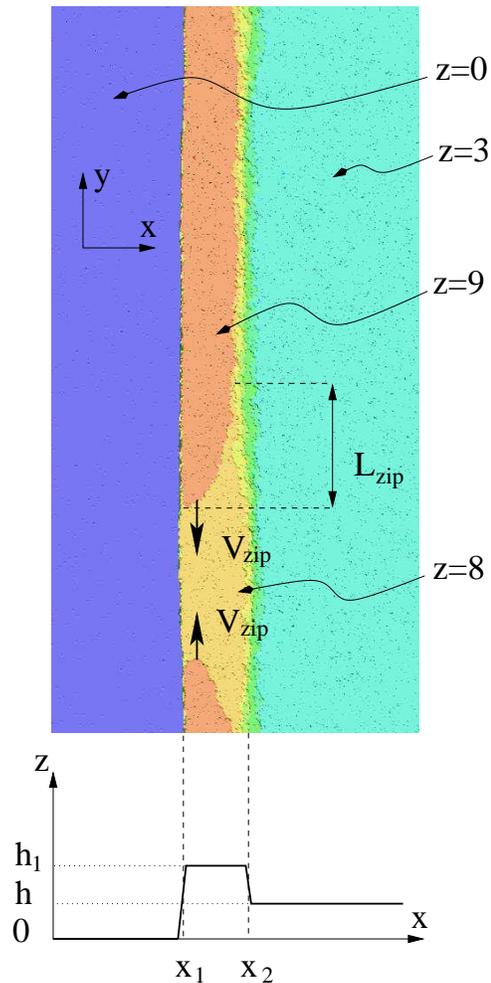}\\
\caption{Zipping of a monolayer island on the rim facet.
Detail of a KMC simulation on a $1000\times1000$ lattice with
$T=0.4$, $E_S=0.5$, $h=3$. On this image, $h_1=8$. }
\label{fig:zipping}
\end{figure}

We now turn to another regime, which appears for
large $E_S$ and small $h$. In this regime, a thick
multilayer rim forms. The top of this rims is facetted.
The increase of the rim height proceeds via the nucleation
of monolayer islands on the rim facet. Once they are nucleated,
these islands grow in size so as to invade the whole rim facet,
as shown on Fig.\ref{fig:zipping}.
Here, we want to focus on the growth process of the
monolayer islands.

\subsection{Monolayer zipping velocity}

Let us consider a straight dewetting front of height $h_1$,
with a rim facet of width $\ell=x_2-x_1$. We aim to analyze the velocity 
of an island zipping along the dewetting front.
We assume that the dewetting front does not deform
and remains straight during this process.
The driving force for the atoms to detach from the front,
and to attach to the monolayer step is the difference between 
the chemical potential of the straight film edge $E_S/h_1$ (see Appendix A),
and the chemical potential of the step $\Omega\tilde\gamma\kappa$, leading to:
\begin{eqnarray}
\Delta\mu={E_S \over h_1}-\Omega \tilde \gamma \kappa,
\end{eqnarray}
where $\kappa$ is the step curvature,
and  $\tilde\gamma$ is the step stiffness.
Here it is assumed that the temperature is 
high enough so that steps properties are isotropic.
As a consequence, we have $\tilde\gamma\approx\gamma$.

The mobility of the atoms for diffusion between the
film edge and the atomic step is $M=Dc_{eq}/(\lambda T)$,
where $\lambda$ is the typical distance between the fronts.
The only available geometric scale for a monolayer rim
zipping along a dewetting front
is the step curvature, and we therefore expect $\kappa\sim 1/\lambda$.

As a summary, the
zipping velocity should take the form:
\begin{eqnarray}
V_{zip}=M\Delta\mu \sim {\Omega D c_{eq}\over T} \kappa 
\left({E_S \over h_1}-\Omega\gamma\kappa\right).
\end{eqnarray}
We assume that the selected curvature 
is the one which maximizes the velocity:
\begin{eqnarray}
\kappa={E_S \over 2h_1\Omega\gamma}.
\end{eqnarray}
Thus, the zipping velocity takes the form
\begin{eqnarray}
V_{zip}\approx C_{zip}{ D c_{eq}  E_S^2 \over  T h_1^2\gamma},
\label{e:Vzip}
\end{eqnarray}
where $C_{zip}$ is an unknown number.

\subsection{Time for the top step to cross the rim facet}
Two-dimensional islands are usually  nucleated
close to the front $h_1$. 
The islands grow in the $y$ direction with velocity $V_{zip}$,
as discussed in the previous subsection. But islands also
grow in the $x$ direction. In order to model the growth
in the $x$ direction we use a 1D model, where a step is parallel
to the front. This 1D model only makes sense
in the limit of small slopes $\partial_yx\ll 1$.
We assume that dynamics is limited by diffusion
on the top of the monolayer island.
The chemical potential difference
between the dewetting front and the
island edge is $E_S/(h_1+1)$. Assuming once again
fast attachment-detachment kinetics, 
the concentration at the step edge and at
the front edge are respectively $c_{eq}$, and  $c_{eq}\exp[E_S/(h_1+1)T]$.
The concentration profile is therefore linear,
and the flux per unit length from the front to the island
edge is:
\begin{eqnarray}
J={D c_{eq} \over x_s}\left({\rm e}^{E_S/(h_1+1)T}-1\right),
\end{eqnarray}
where $x_s$ is the distance between  the dewetting
front and the island edge.
The velocity of the island edge in the $x$ direction
is therefore: $dx_s/dt=\Omega J$. From this equation,
we find 
\begin{eqnarray}
x_s=\left[2\Omega D c_{eq}\left( {\rm e}^{E_S/(h_1+1)T}-1\right)t\right]^{1/2}.
\end{eqnarray}
The step will reach the step bunch at $x_s=\ell$ in a time $t_\ell$.
For weak driving forces, $E_S\ll Th_1$, one has
\begin{eqnarray}
t_\ell={\ell^2 \over 2\Omega D c_{eq}}
(h_1+1) {T \over E_S}.
\label{e:tell}
\end{eqnarray}

\subsection{Zipping length}

The direct observation of $V_{zip}$ or $t_\ell$ in the
KMC simulations would require delicate front tracking
procedures and we shall look for a quantity which
can be observed on a single snapshot.
From  relation (\ref{e:tell}),
the tip length of the zipping monolayer reads:
\begin{eqnarray}
L_{zip}=V_{zip}t_\ell\approx C_{zip} {E_S \over 2\Omega \gamma}
\ell^2
\,{h_1+1\over h_1^2}.
\label{e:Lzip}
\end{eqnarray}
The length $L_{zip}$ is the length between the tip of the 
zipping monolayer (in the vicinity of the 
front of height $h_1$), and the point where the step edge
of this monolayer reaches the bunch of height $h_2$
(the distance is measured along $y$).

Since the nucleation barrier for the formation
of new monolayer islands grows with $h_1$, the typical distance
between island nucleation sites increases exponentially with
$h_1$, as shown in Ref.\cite{PierreLouis2009}. Thus,
the length $L_{zip}$ is always smaller than the typical
distance between island nucleation sites. Therefore,
for large enough $h_1$, the islands first reach the total
width $\ell$ of the facet, and then zip along the dewetting
front with a tip length $L_{zip}$. This is indeed the scenario
observed in KMC simulations.

\subsection{Comparison to KMC}

We have performed KMC simulations, with $E_S=0.5$, $T=0.4$, $h=3$,
in a $1000\times 1000$ system with total simulation 
time $t=3.5\times 10^8$. The simulation time is chosen to be large
enough for the system to reach the late stage regime where 
$L_{zip}\gg\ell$. Measurements were performed
on five KMC simulations snapshots,
each time measuring $L_{zip}$ from an average
of its value at the two ends of the expanding island on the top facet.
From each measurement, we have extracted the value of
$C_{zip}$ from Eq.(\ref{e:Lzip}) with $\gamma=0.42$\cite{Krishnamachari1996,PierreLouis2007,PierreLouis2009}.
The results are reported in Table I.
We obtain the average value
\begin{eqnarray}
C_{zip}\approx 0.25\pm 0.05.
\label{e:Czip}
\end{eqnarray}
The error here  provides an indication of the order
of magnitude for the observed variations 
of $C_{zip}$ extracted from different measurements.
The result (\ref{e:Czip}) does not include the case where $h_1=6$,
for which the condition $\ell\ll L_{zip}$ is not verified.
Nevertheless, we see in Table \ref{table:1} that the
results with $h_1=6$ would provide similar values for $C_{zip}$.

\begin{table}
\begin{tabular}{|c|c|c|c|}
  \hline
  $h_1$ & $\ell$ & $L_{zip}$  & $C_{zip}$ \\
  \hline
  6 & 56 & 86  & 0.24 \\
  6 & 56 & 54  & 0.15 \\
  7 & 69 & 150 & 0.31 \\
  7 & 85 & 180 & 0.24 \\
  8 & 83 & 124 & 0.20 \\
  \hline
\end{tabular}
\caption{Observed values of $\ell$ and $L_{zip}$ from KMC simulations
with $T=0.4$, $E_S=0.5$, and $h=3$ in a $1000\times1000$ lattice.}
\label{table:1}
\end{table}

Finally, we would like to point out that we have not yet studied systematically
all possible regimes of zipping, due to limitations in the
our KMC simulations. Indeed, for smaller
values of $E_S$, the simulation time is too large,
and for larger values of $E_S$, the solid on solid
restriction (which forbids overhangs) should be released.
Important improvements in our numerical approach
are therefore needed in order to address these questions.

\section{Conclusion}
In conclusion, we have analyzed the motion of atomic steps during the 
dewetting of ultra-thin solid films with various geometries: 
(i) simultaneous growth of a monolayer island with a hole; 
(ii) expansion of a hole with a monolayer rim;
(iii) zipping of a monolayer island on the top facet of a thick rim. 

Our KMC results are in good agreement
with step models where mass transport is limited by diffusion on terraces
between steps. These results demonstrate that the
step model approach provides a systematic way
to investigate the evolution dynamics of thin films
during the dewetting process. 

Several additional
ingredients could be added to the step models, such as
anisotropy, non-trivial attachment-detachment kinetics
(e.g. an Ehrlich-Schwoebel barrier, or step transparency),
or elastic interactions between steps. We hope to
report along these lines in the future.

We acknowledge support from "nanomorphog\'en\`ese"
and "D\'eFiS" ANR-PNANO grants.

\begin{appendix}
\section{Local chemical potential at the edge of a facetted film}
\label{a:chem_pot_edge_hole}
The total free energy of a film of height $h_0$ at $x>x_0(y)$
(the height is zero at $x<x_0(y)$) is:
\begin{eqnarray}
{\cal F}=\int ds_0 \;\gamma_{edge}- E_S \Omega^{-1}\int dy \;x_0(y),
\end{eqnarray}
where $s_0$ is the arclength along the edge of the film,
and $\gamma_{edge}$ is the free energy per unit length of the 
edge of the film. The total number of atoms in the film is:
\begin{eqnarray}
{\cal N}={\cal N}_0-\Omega^{-1}h_0\int dy \;x_0(y),
\end{eqnarray}
where ${\cal N}_0$ is a constant.
Equilibrium reads:
\begin{eqnarray}
\delta\left({\cal F}-\mu{\cal N}\right)=0.
\end{eqnarray}
Our variable here is $x_0(y)$, the position of the edge.
We therefore have:
\begin{eqnarray}
\mu={\delta{\cal F}\over \delta x_0(y)}\left(\delta{\cal N}\over \delta x_0(y)\right)^{-1}.
\end{eqnarray}
Using the above expressions:
\begin{eqnarray}
\mu=\Omega \tilde \gamma_0 \kappa_0 +  {E_S \over h_0},
\end{eqnarray}
where $\tilde\gamma_0=(\gamma_{edge}+\gamma_{edge}'')/h_0$  is the stiffness
per layer of the film edge (If the edge 
is a bunch of steps, and if the step-step interaction
is negligible, then $\tilde\gamma_0$ is the step stiffness).
Note that this derivation of the chemical potential
relies on the postulate that the film profile across
the film edge does not vary along along the film edge.

\end{appendix}

\bibliography{references}

\end{document}